\documentclass[reprint, amsmath,amssymb, aps, pra, longbibliography]{revtex4-1}

\usepackage{soul}
\usepackage{amsmath}   
\usepackage{amssymb}
\usepackage{graphicx}
\usepackage{dcolumn}
\usepackage{bm}
\usepackage{amsfonts}
\usepackage{subfigure}
\usepackage{array}
\usepackage{float}
\usepackage{color}
\usepackage[colorlinks=true,linkcolor=blue]{hyperref}%
\hypersetup{allcolors=blue}
\usepackage[normalem]{ulem}
\usepackage{xcolor}

\newcommand{\ket}[1]{\ensuremath{\left\vert #1 \right\rangle}}

\newcommand{\gfix}[1]{\textcolor{black}{#1}}

\begin{document}

\title{Hyperfine Structure of $nP_{1/2}$ Rydberg States in $^{85}$Rb}
\date{\today }

\author{R. Cardman}
\email{rcardman@umich.edu}
\affiliation{Department of Physics, University of Michigan, Ann Arbor, MI 48109, USA}
\author{G. Raithel}
\affiliation{Department of Physics, University of Michigan, Ann Arbor, MI 48109, USA}

\begin{abstract}
We measure the hyperfine structure of $nP_{1/2}$ Rydberg states using mm-wave spectroscopy on an ensemble of laser-cooled $^{85}$Rb atoms. Systematic uncertainties in our measurement from the Zeeman splittings induced by stray magnetic fields and dipole-dipole interactions between two Rydberg atoms are factored in with the obtained statistical uncertainty. Our final measurement of the $nP_{1/2}$ hyperfine coupling constant is $A_{\text{hfs}}=1.443(31)$~GHz. This measurement is useful for studies of long-range Rydberg molecules, Rydberg electrometry, and quantum simulation with dipole-dipole interactions involving $nP_{1/2}$ atoms. 
\end{abstract}

\maketitle

%\section{Introduction}
The Rydberg $nP_{j}$ states of Rb afford the capability of studying long-range molecular interactions in macrodimers~\cite{Boisseau2002,Hollerith2019,Hollerith2021}, Rydberg-ground pairs~\cite{Niederpruem2016a, Niederpruem2016b}, and, recently predicted and observed, Rydberg-ion mixtures~\cite{Duspayev2021, Deiss2021, Zuber2022}. Furthermore, their couplings with states of different parities are useful in understanding dipole-dipole interactions~\cite{Li2005,Ravets2014} and employing quantum electrometry of resonant rf waves via Autler-Townes splitting, observable through electromagnetically-induced-transparency (EIT) spectroscopy~\cite{Holloway2014, Anderson2022}. Hyperfine interactions of the nuclear magnetic moment and electric quadrupole moment with the angular momentum of the valence electron typically are not observable in $nP_{j}$ Rydberg states through laser-based spectroscopic methods due to limitations in frequency resolution (energy splittings are on the order of kHz), although hyperfine effects have been experimentally presented in Cs~\cite{Ripka2022}. Millimeter-wave spectroscopy of Rydberg molecular states involving $nP_{j}$ atoms could provide insights in the role of hyperfine coupling on the adiabatic potentials of the molecules, for the spectroscopic measurement is, in principle, only limited by the Rydberg-state lifetime and the rf-field interaction time. As a consequence, knowledge of the hyperfine structure (HFS) is essential for predicting these quantum behaviors.

\par While precision measurements for the hyperfine-coupling constants in $nP$ Rydberg states have been provided before in~\cite{Belin1974,Belin1976a,Belin1976b,Farley1977} on $^{133}$Cs and $^{87}$Rb for $n<13$, no coupling constant has been provided for $^{85}$Rb using principal quantum numbers greater than $n=8$, at which point the hyperfine interaction does not scale with $n$. In~\cite{Li2003}, the $nP_{1/2}$ HFS is observable for both $^{85}$Rb and $^{87}$Rb (see their Fig. 2). However, the thermal atomic beam used contributed to a significant amount of Doppler broadening, and a measurement was not provided. 
\par In the present work, we perform mm-wave resonance spectroscopy in the Ka- and U-bands on ultracold $^{85}$Rb. Thus, in the absence of any Doppler and transit-time broadening, we obtain Fourier-limited spectral lines of the \gfix{$\ket{nS_{1/2}, F=3, m_F}\rightarrow \ket{nP_{1/2}, F', m_{F'}}$} transitions and use the splitting between the $F'=2$ and $F'=3$ \gfix{hyperfine} peaks in order to arrive at an $n$-independent, HFS-coupling-constant $A_{\text{hfs}}$ measurement for $nP_{1/2}$ Rydberg states. The spectroscopic series involves $n=42-44$ and $46$. \gfix{Careful} cancellation of stray magnetic fields to $< 5$~mG is necessary to observe symmetric, Fourier-limited spectral features for both peaks. Our uncertainty budget, as a result, takes into account the role of the background magnetic field on our measurement. Additionally, we provide a systematic uncertainty arising from electric dipole-dipole interactions between $nS_{1/2}$ and $nP_{1/2}$ atoms. 
%\section{Theory}
\par An alkali metal like $^{85}$Rb features a single valence electron of total angular momentum $\textbf{J}$, spin $\textbf{S}$, and orbital angular momentum $\textbf{L}$. The nucleus of the given isotope features an intrinsic angular momentum $\textbf{I}$ associated with the net magnetic moments of all contained nucleons. For $^{85}$Rb, the nuclear spin quantum number is $I=5/2$. In general, the hyperfine shift \gfix{of a $nP_{j}$ level with hyperfine quantum number $F'$} is, in atomic units, 
\begin{multline}
    \Delta_{\text{hfs}} = \frac{A_{\text{hfs}}}{[n-\delta_{lj}(n)]^3}\langle \textbf{I}\cdot\textbf{J}\rangle\\
    + \frac{B_{\text{hfs}}}{[n-\delta_{lj}(n)]^3}\bigg<\frac{3(\textbf{I}\cdot\textbf{J})^2+\frac{3}{2}\textbf{I}\cdot\textbf{J}-IJ(I+1)(J+1)}{2IJ(2I-1)(2J-1)}\bigg>,
\end{multline}
where $\delta_{lj}(n)$ is the $nlj$-dependent quantum defect~\cite{Li2003}, the first term describes the magnetic dipole-dipole interaction between the nucleus and Rydberg electron, and the second term quantifies the nuclear electric-quadrupole interaction. A third term, immeasurable in this type of experiment, involves magnetic-octupole interactions between the two particles~\cite{SteckRb}. For $nP_{1/2}$ states, only $A_{\text{hfs}}$ is nonzero. 
\par \gfix{Due to the large extent of the Rydberg electron wave function, short-range interactions scale as $[n-\delta_{lj}(n)]^{-3}$~\cite{GallagherBook}.} Thus, the measured splitting $\nu_{\text{hfs}}$ between $F'=2$ and $F'=3$ can be expressed as \begin{equation}
    \nu_{\text{hfs}}=\frac{3A_{\text{hfs}}}{[n-\delta_{lj}(n)]^3},
    \label{eq:HFSFormula}
\end{equation}
where the units of $A_{\text{hfs}}$ are GHz. 
%\section{Experimental Methods}
\par
\begin{figure}[htb]
 \centering
  \includegraphics[width=0.45\textwidth]{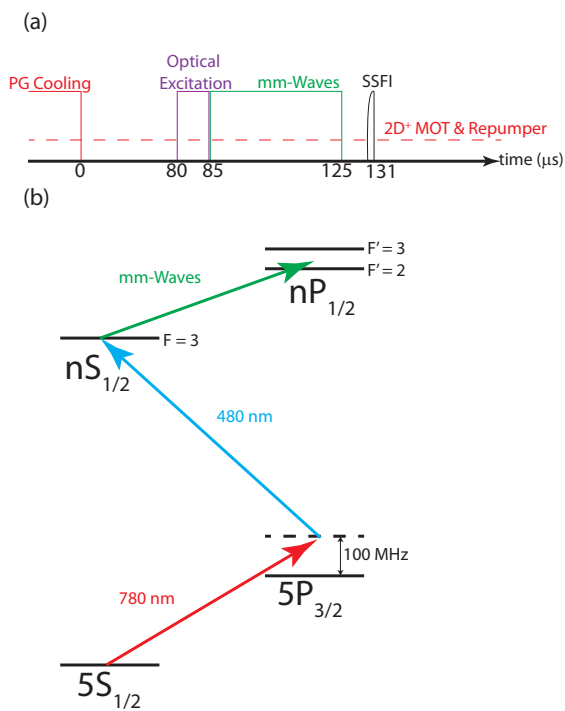}
  \caption{(Color online) Timing sequence of an experimental cycle is shown in (a). "Optical excitation" refers to the simultaneous 780-nm and 480-nm pulses. The 2D$^+$ MOT and repumping laser beams are always on. In (b), we show the level diagram of $^{85}$Rb states relevant to the experiment (not drawn to scale). Atoms are excited off-resonantly from the upper hyperfine level of the $5S_{1/2}$ state into the $nS_{1/2}$ Rydberg state during the "optical excitation" pulse. There is a statistical mixture of $F=2$ and $F=3$ Rydberg states after the optical excitation, but the number of atoms in the $F=2$ states is too small to achieve an appropriate signal-to-noise ratio during the spectroscopic mm-wave pulse. Thus, the mm-wave frequency scan range is set to only probe the atoms in $F=3$ state.} 
  \label{fig:timingandlevel}
\end{figure}

  In our experiment, a slow atomic beam of $^{85}$Rb prepared by a  continuously operating 2D$^{+}$ MOT~\cite{Dieckmann1998} is captured and cooled via polarization gradients (PG) in the $\sigma^{+}$-$\sigma^{-}$ configuration~\cite{Dalibard1989} for $14.2~$ms. We leave the 2D$^{+}$ MOT laser beams and all repumping beams on throughout the duration of the experiment. The D2-molasses cooling light is switched off for 80~$\mu$s before $5-\mu$s-long  optical excitation beams are switched on. These beams produce $nS_{1/2}$ Rydberg atoms used for the mm-wave spectroscopy, where $n=42$-44 and 46. A $40-\mu$s mm-wave pulse drives \gfix{the $\ket{nS_{1/2},F=3, m_F}\rightarrow\ket{nP_{1/2},F'=2 \, {\rm{or}} \, 3, m_F}$} transitions necessary for determining the hyperfine splitting. At the end of the mm-wave exposure time, an electric field is smoothly ramped up to $100-150$~V/cm in $1~\mu$s for state-selective field ionization (SSFI) of the $nS_{1/2}$ and $nP_{1/2}$ levels~\cite{GallagherBook}. $^{85}$Rb$^{+}$ counts are detected with a micro-channel-plate detector (MCP). A timing sequence for the experimental cycle is given in Fig.~\ref{fig:timingandlevel}(a). 

\par Optical excitation from the upper hyperfine level of the ground state is provided in the form of an off-resonant, two-photon transition using $780$- and $480$-nm pulses, described in the quantum-state diagram of Fig.~\ref{fig:timingandlevel}(b). A 780-nm external-cavity diode laser (ECDL) is tuned 100~MHz above the upper-most hyperfine level of the $5P_{3/2}$ state, while a 960-nm ECDL, amplified and doubled to make $480$-nm light, is tuned to make up the resonance with the Rydberg state. Polarizations of the optical excitation beams and atomic sample, as well as the blue-detuning of the 780-nm laser from the upper-most hyperfine level of the $5P_{3/2}$ state result in significantly more Rydberg-atom population in \gfix{$\ket{nS_{1/2},F=3, m_F}$ than $\ket{nS_{1/2},F=2, m_F}$}. Therefore, we perform our mm-wave spectroscopy only on the $F=3$ hyperfine levels for all $n$ studied.

\begin{figure}[htb]
 \centering
  \includegraphics[width=0.53\textwidth]{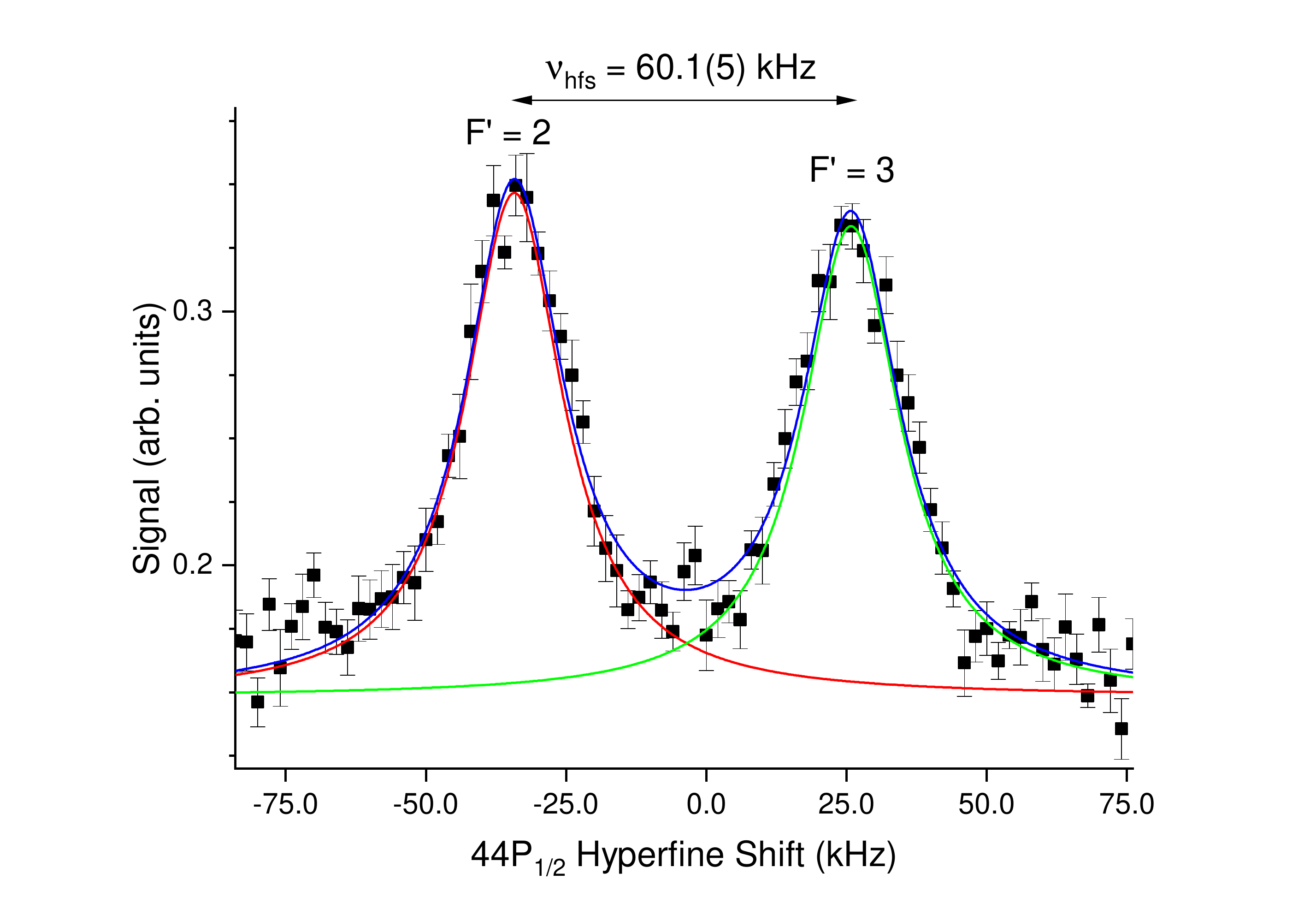}
  \caption{Single-photon resonance spectrum of the $\ket{44S_{1/2},F=3}\rightarrow\ket{44P_{1/2},F'}$ transition using mm-waves. The spectrum shown is an arithmetic mean of 8 individual spectra. Each individual spectrum is averaged over 400 experimental cycles. On the frequency axis, we show the $44P_{1/2}$ hyperfine shifts for each $F'$ level with respect to the center-of-gravity transition frequency, $\nu_{0}=45.113624~$GHz, i.e., the frequency of the \gfix{$\ket{44S_{1/2},F=3, m_F}\rightarrow\ket{44P_{1/2}}$ } transition with the $44P_{1/2}$ hyperfine structure removed. Each scatter point corresponds to a frequency step size of 2~kHz. Signal error bars for the scatter points indicate the standard error of the mean (SEM) over the 8 individual points acquired. In this spectrum, the total detected count rate is below two ions per experimental cycle. The solid curves are the double- and individual-Lorentzian fit functions from which the peak centers are acquired to measure the HFS splittings. Measured linewidths are $21(1)~$kHz for both peaks. } 
  \label{fig:spectrum}
\end{figure}

\begin{figure*}[t]
 \centering
  \includegraphics[width=0.6\textwidth]{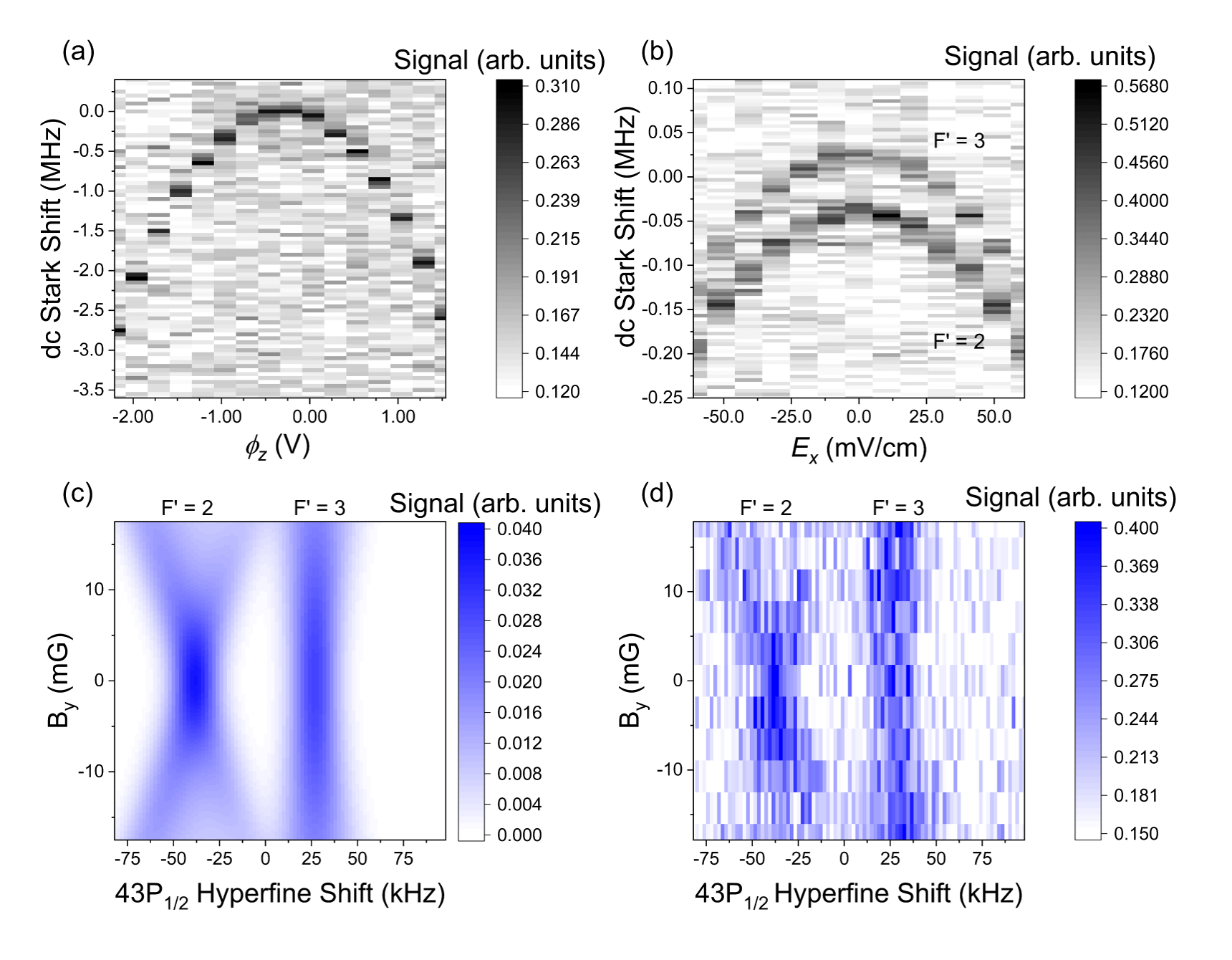}
  \caption{ This figure explains the static electromagnetic field zeroing process necessary for a HFS measurement. In (a), we show a map of dc Stark shifts on the $\ket{44S_{1/2}}\rightarrow\ket{44P_{1/2}}$ transition as a function of applied potential $\phi_{z}$ on plate electrodes in order to find the voltage that cancels shifts from stray electric fields along the $z$-direction. The differential dc polarizability between the two states is $\alpha_{44P_{1/2}}-\alpha_{44S_{1/2}} = 9.564$~kHz/(V/m)$^{2}$. Here, the mm-wave frequency steps are not resolved enough to observe the HFS splitting. (b) Verification that the hyperfine splitting is not affected by electric fields smaller than $60~$mV/cm. The $\ket{43S_{1/2}, F=3}\rightarrow\ket{43P_{1/2}, F'}$ spectra are plotted as a function of applied electric field $E_x$ in the $x$-direction with stray fields canceled in the other two directions.  (c) Calculated Zeeman splitting of the $\ket{43S_{1/2},F=3}\rightarrow\ket{43P_{1/2},F'}$ transitions for $F'=2$ and 3. This calculation is for the case that the applied magnetic field is perpendicular ($y$-direction) to the mm-wave polarization. In (d), we show an experimental analogue to our calculation.  } 
  \label{fig:HFSPaperFig3}
\end{figure*}

\par The mm-waves are synthesized by an Agilent MXG Analog Signal Generator (Model N5183A) that is referenced to an SRS Model FS725 Rubidium Frequency Standard. For spectroscopy of $n=42$-44, the synthesized mm-waves are frequency doubled by a SAGE Model SFA-192KF-S1 active X2 multiplier and broadcast from $\simeq$40~cm to the $nS_{1/2}$ Rydberg atoms with a horn antenna. We do not double the mm-waves at $39.121~$GHz for the $\ket{46S_{1/2},F=3, m_F}\rightarrow\ket{46P_{1/2},F', m_F}$ spectrum and directly connect the synthesizer to a standard-gain horn antenna, located $\simeq$30~cm from the spectroscopic interaction region, with a 20-dBi directivity. 
%\section{Results}

\par Spectra of the $\ket{nS_{1/2},F = 3, m_F}\rightarrow\ket{nP_{1/2},F', m_F}$ transition were acquired for each $n$ in the $n=42-44$ and 46 series. A double Lorentzian was fit to an arithmetic average of eight experimental scans of the mm-wave frequency over the two hyperfine lines. In order to determine $\nu_{\text{hfs}}$ from our data, we take the difference between the line centers of the Lorentzian fit functions. The uncertainties in the line centers were added in quadrature and used as the uncertainty in the HFS splitting, $\delta\nu_{\text{hfs}}$. Fig.~\ref{fig:spectrum} shows a typical spectrum obtained in this study. The linewidths are at the level of the Fourier limit ($0.89/40~\mu$s$=22~$kHz), meaning the Rabi frequencies of the transitions are in the range of $10~$kHz, preventing any observable ac Stark shifts.

Once we obtain $\nu_{\text{hfs}}$, we use the $\delta_{0}$ and $\delta_{2}$ quantum-defect values for Rb $nP_{1/2}$ measured in~\cite{Li2003} and the Rydberg-Ritz equation~\cite{GallagherBook} to derive a measurement for $A_{\text{hfs}}$ using Eq.~\ref{eq:HFSFormula}. These quantities are $\delta_{0}=2.6548849(10)$ and $\delta_{2}=0.2900(6)$.  Because the uncertainties in $\delta_{0}$ and $\delta_{2}$ lead to shifts much smaller than our measurement uncertainties, we neglect them in our uncertainty budget. Thus, $\delta A_{\text{hfs}}/A_{\text{hfs}}=\delta\nu_{\text{hfs}}/\nu_{\text{hfs}}$. Table~\ref{measurements} lists $\nu_{\text{hfs}}$ and $A_{\text{hfs}}$ for a given $n$ in the range $n = 42-44$ and 46. A weighted average and uncertainty over all $n$ provides a final value for $A_{\text{hfs}}$ and a statistical uncertainty, also included in the table.

\begin{table}[htbp]
\caption{\label{measurements} Summary of HFS splittings and derived $A_{\text{hfs}}$ using Eq.~\ref{eq:HFSFormula} and $\delta_{0}=2.6548849(10)$,~$\delta_{2}=0.2900(6)$~\cite{Li2003}. }
\begin{ruledtabular}
\begin{tabular}{c|c|c}
    $n$ & $\nu_{\text{hfs}}$~(kHz) & $A_{\text{hfs}}$~(GHz)\\
    \hline
   42 &  72.7(6) & 1.476(12) \\ 
    43 & 65.3(6) &  1.429(13)\\
    44 & 60.1(5) & 1.416(12)\\
   46 & 54(1) & 1.466(27) \\
\end{tabular}
\begin{tabular}{c|c}
     $A_{\text{hfs}}$, weighted average~(GHz) & 1.443 \\
    Statistical uncertainty~(GHz)&   0.007  \\
\end{tabular}
\end{ruledtabular}
\end{table}

\par Symmetry of our observed spectral lines indicates that background electric- and magnetic-field inhomogeneities are negligible. A set of six, orthogonal plate electrodes situated in our science chamber is used to cancel electric fields below 50~mV/cm by observing shifts in $\ket{nS_{1/2}}\rightarrow\ket{nP_{1/2}}$ spectra as a function of applied electric field; a map of these spectra is shown along the $z$-axis in Fig.~\ref{fig:HFSPaperFig3}(a) for $n=44$. Fig.~\ref{fig:HFSPaperFig3}(b) displays a more resolved map for $n=43$ with an applied field along the $x$-direction. Electric fields contribute no systematic shift in the HFS splitting because the $nP_{1/2}$ Rydberg states lack a tensor polarizability that would otherwise cause distortions in the $F'=2$ and $F'=3$ peaks as a result of $|m_{F'}|$ splititngs. Therefore, both $F'$ states and all $|m_{F'}|$ undergo the same dc Stark shifts leaving $\nu_{\text{hfs}}$ insensitive to stray electric fields. This insensitivity is verified in Fig.~\ref{fig:HFSPaperFig3}(b) for $n=43$, where we apply electric field $E_{x}$ magnitudes up to $60~$mV/cm and scan over both hyperfine peaks of the $43P_{1/2}$ state. Inhomogeneous broadening from position-dependent electric fields within the atom cloud is the only possible dc Stark effect, which is negligible as exhibited by the line symmetries and linewidths near the Fourier limit of $22$~kHz. Excessive magnetic fields within the interaction region on the other hand do distort measurements of $\nu_{\text{hfs}}$ from the Zeeman splittings of the $m_F$ and $m_{F'}$ sublevels, as seen in the following paragraph. 

\begin{figure}[htb]
 \centering
  \includegraphics[width=0.5\textwidth]{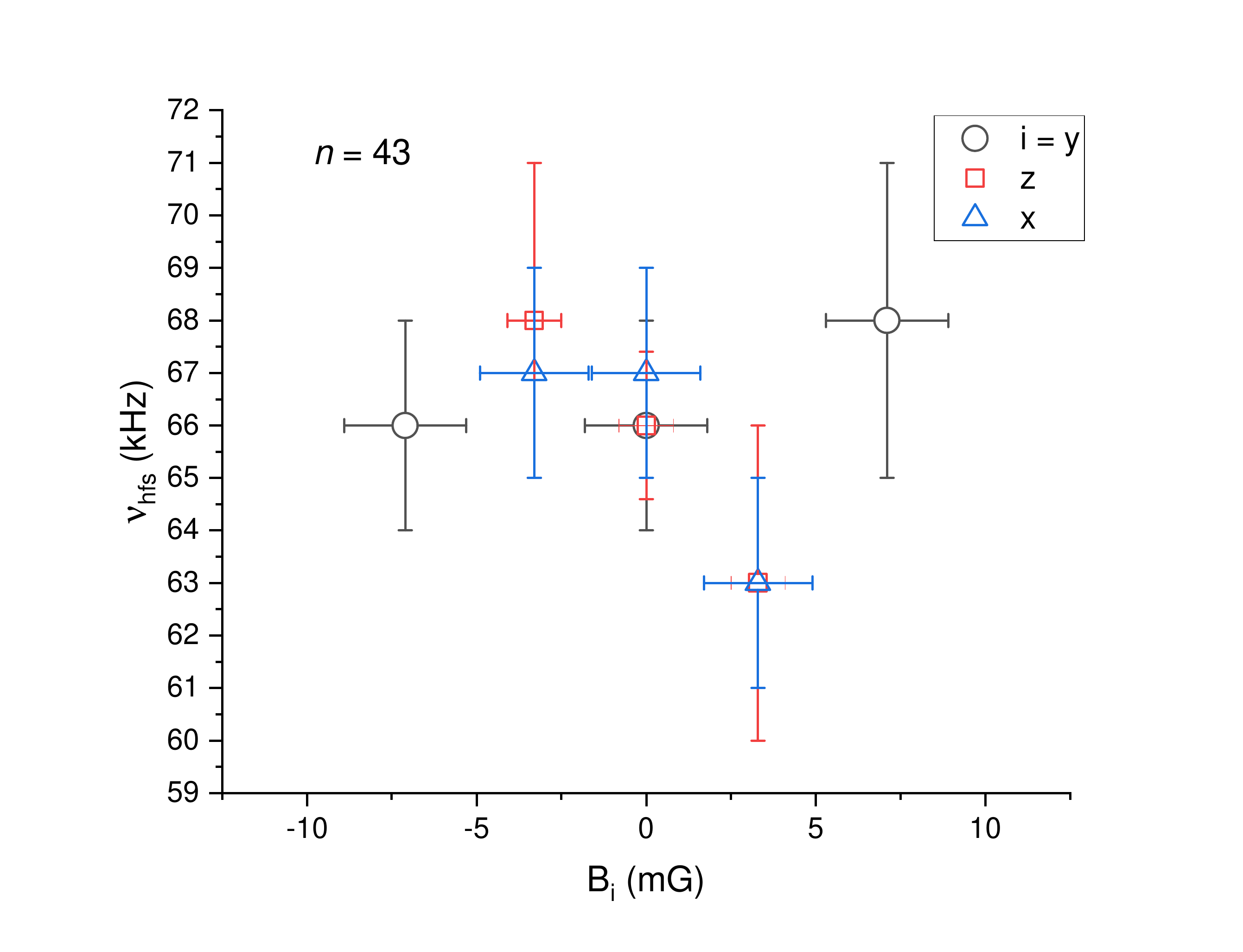}
  \caption{ Measured $\nu_{\text{hfs}}$ of $43P_{1/2}$ for given applied magnetic fields in all three spatial dimensions. The SEM of all nine $\nu_{\text{hfs}}$ is used as the systematic $\delta\nu_{\text{hfs}}$ from potential stray magnetic fields. Magnetic field uncertainties arise from noise in our current sources.} 
  \label{fig:magneticfield}
\end{figure}

\par Three pairs of externally located Helmholtz coils apply homogeneous magnetic fields to eliminate Zeeman broadening and splitting of the $F'$ states~\cite{Ramos2019}. Expected and observed behaviors of the Zeeman splittings for $n=43$ are shown in Figs.~\ref{fig:HFSPaperFig3}(c) and (d), respectively, for the case of a magnetic field perpendicular to the mm-wave polarization. Our stray magnetic fields are reduced down to a magnitude no greater than $5$~mG. In order to quantify the possible systematic uncertainties from any leakage within this range, we take the standard error of the mean (SEM) in a sample of splittings at $n=43$ by offsetting our compensation magnetic fields within 10~mG of the cancellation values in all three directions $x,~y,$~\&~$z$ independently. A similar analysis was done in the context of measuring the $nS_{1/2}$ HFS for $^{85}$Rb Rydberg states~\cite{Ramos2019}. This distribution is presented in Fig.~\ref{fig:magneticfield}. Our SEM yields $\delta\nu_{\text{hfs}}= 0.6$~kHz at $n=43$ and $\delta A_{\text{hfs}}=13$~MHz.

\par
\begin{figure}[htb]
 \centering
  \includegraphics[width=0.53\textwidth]{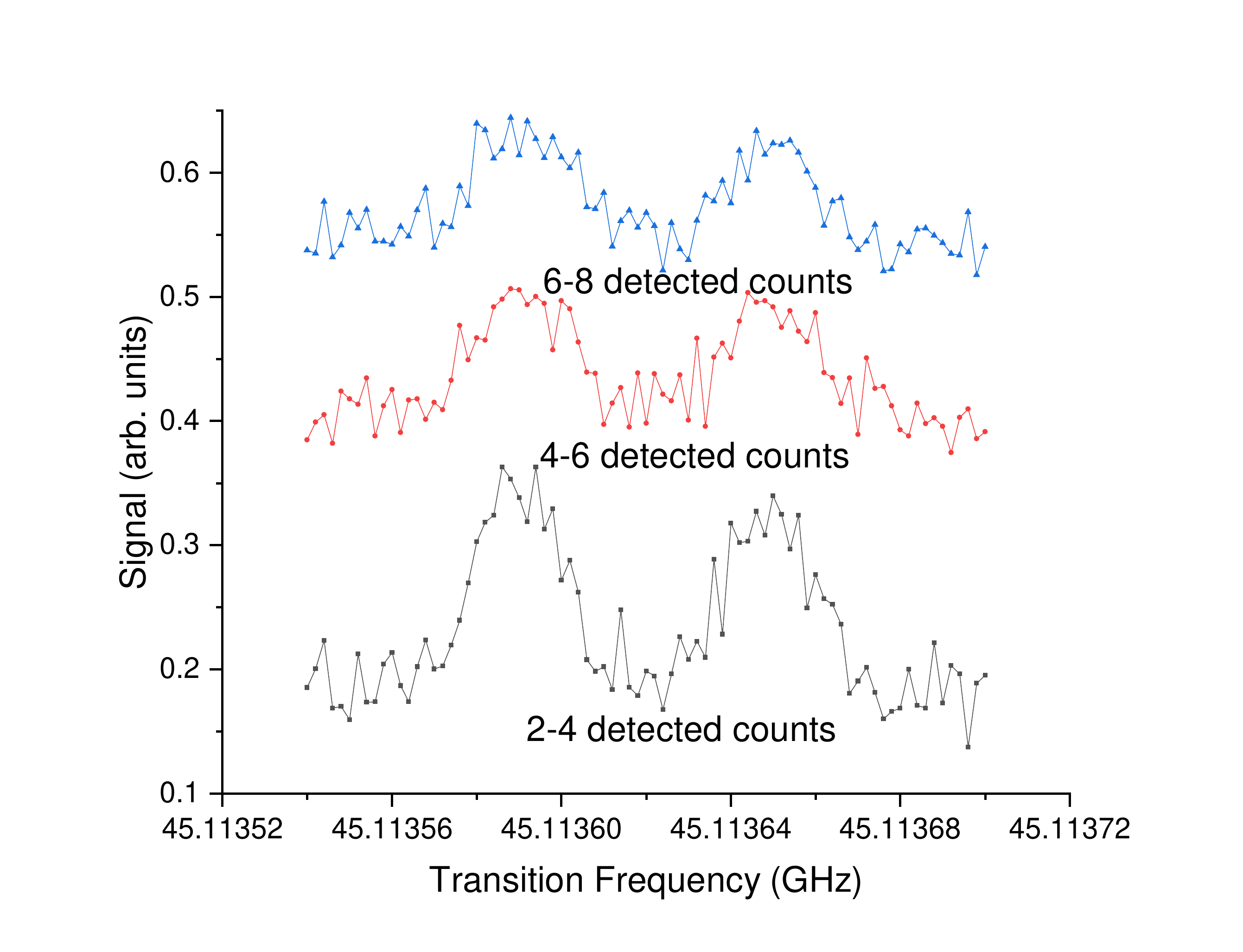}
  \caption{In this series, the $44P_{1/2}$ HFS is measured for three different bins of total detected ion counts from field-ionized Rydberg atoms. The increase in count rate is achieved by starting the optical excitation 5-10~$\mu$s earlier to prolong the laser pulse duration. Because the atomic density is rising proportionally with the count rate, the dipole-dipole shifts, if they are significant, should increase as well. There is no apparent dipole-dipole shift over 1~kHz, implying that the atomic spacing must be at a minimum of $120~\mu$m. } 
  \label{fig:counts}
\end{figure}

 We also take into account shifts from dipole-dipole interactions between one atom with an internal state of $nS_{1/2}$ and another with that of $nP_{1/2}$. In Fig.~\ref{fig:counts}, we exhibit that the shift in $\nu_{\text{hfs}}$ does not exceed $1$~kHz for $n=44P_{1/2}$, as the maximum ion-count rate, and therefore density, is increased by a factor four by prolonging the optical excitation time up to $15$~$\mu$s. All measurements in Table~\ref{measurements} were taken with fewer than three detected total counts. An upper-limit of the $C_{3}$ coefficient is estimated to be $1.7~$GHz~$\mu$m$^{3}$ for $n=44$ by finding and fitting adiabatic potentials of Rydberg-Rydberg molecules~\cite{Han2018}. This estimate implies that the atomic spacing is $R\gtrsim 120~\mu$m and the systematic uncertainty in $A_{\text{hfs}}$ from dipole-dipole interactions has an upper limit of $27$~MHz. Higher-order Rydberg-Rydberg interactions, such as van der Waals shifts between two atoms of the same internal state are at the order of 1~mHz or less for these $n$ and therefore are not included in our overall uncertainty budget~\cite{Ramos2019}.

\par We present our uncertainty budget in Table~\ref{uncertainties}. Adding the three sources in quadrature, we find the overall uncertainty to be $\delta A_{\text{hfs}}=31$~MHz. 

%\section{Discussion}
\begin{table}[htbp]
\caption{\label{uncertainties} Uncertainty budget for a measurement of $A_{\text{hfs}}$. }
\begin{ruledtabular}
\begin{tabular}{c|c}
    Source & $\delta A_{\text{hfs}}$ (GHz)\\
\hline
     Dipole-dipole interactions & 0.027 \\
      Stray magnetic fields & 0.013 \\
    Statistical uncertainty &   0.007  \\
\end{tabular}
\end{ruledtabular}
\end{table}

%\section{Conclusion}
\par In summary, we measured the hyperfine coupling constant $A_{\text{hfs}}$ for Rydberg-$nP_{1/2}$ states of $^{85}$Rb using mm-wave spectroscopy with Fourier-limited linewidths. Our precision in $A_{\text{hfs}}$ is mainly limited by the estimated lower limit of  atomic spacing within our Rydberg cloud that may lead to dipole-dipole interactions. In addition to our measurement's applicability for investigating ultracold Rydberg chemistry~\cite{Boisseau2002,Niederpruem2016a,Niederpruem2016b,Hollerith2019,Hollerith2021,Duspayev2021,Deiss2021,Zuber2022} and dynamic electric-field sensing of rf waves with thermal Rydberg atoms~\cite{Holloway2014,Anderson2022,Ripka2022}, the HFS of $nP_{1/2}$ states can possibly be included in models for quantum simulation~\cite{Khazali2022}.

\section*{ACKNOWLEDGMENTS}
This work was supported by NSF Grants No. PHY-2110049 and PHY-1806809 and NASA Grant No. NNH13ZTT002N. R.C. acknowledges support from the Rackham Predoctoral Fellowship. 

\bibliographystyle{apsrev4-1}
\bibliography{References.bib}

\begin{thebibliography}{25}%
\makeatletter
\providecommand \@ifxundefined [1]{%
 \@ifx{#1\undefined}
}%
\providecommand \@ifnum [1]{%
 \ifnum #1\expandafter \@firstoftwo
 \else \expandafter \@secondoftwo
 \fi
}%
\providecommand \@ifx [1]{%
 \ifx #1\expandafter \@firstoftwo
 \else \expandafter \@secondoftwo
 \fi
}%
\providecommand \natexlab [1]{#1}%
\providecommand \enquote  [1]{``#1''}%
\providecommand \bibnamefont  [1]{#1}%
\providecommand \bibfnamefont [1]{#1}%
\providecommand \citenamefont [1]{#1}%
\providecommand \href@noop [0]{\@secondoftwo}%
\providecommand \href [0]{\begingroup \@sanitize@url \@href}%
\providecommand \@href[1]{\@@startlink{#1}\@@href}%
\providecommand \@@href[1]{\endgroup#1\@@endlink}%
\providecommand \@sanitize@url [0]{\catcode `\\12\catcode `\$12\catcode
  `\&12\catcode `\#12\catcode `\^12\catcode `\_12\catcode `\%12\relax}%
\providecommand \@@startlink[1]{}%
\providecommand \@@endlink[0]{}%
\providecommand \url  [0]{\begingroup\@sanitize@url \@url }%
\providecommand \@url [1]{\endgroup\@href {#1}{\urlprefix }}%
\providecommand \urlprefix  [0]{URL }%
\providecommand \Eprint [0]{\href }%
\providecommand \doibase [0]{http://dx.doi.org/}%
\providecommand \selectlanguage [0]{\@gobble}%
\providecommand \bibinfo  [0]{\@secondoftwo}%
\providecommand \bibfield  [0]{\@secondoftwo}%
\providecommand \translation [1]{[#1]}%
\providecommand \BibitemOpen [0]{}%
\providecommand \bibitemStop [0]{}%
\providecommand \bibitemNoStop [0]{.\EOS\space}%
\providecommand \EOS [0]{\spacefactor3000\relax}%
\providecommand \BibitemShut  [1]{\csname bibitem#1\endcsname}%
\let\auto@bib@innerbib\@empty
%</preamble>
\bibitem [{\citenamefont {Boisseau}\ \emph {et~al.}(2002)\citenamefont
  {Boisseau}, \citenamefont {Simbotin},\ and\ \citenamefont
  {C\^ot\'e}}]{Boisseau2002}%
  \BibitemOpen
  \bibfield  {author} {\bibinfo {author} {\bibfnamefont {C.}~\bibnamefont
  {Boisseau}}, \bibinfo {author} {\bibfnamefont {I.}~\bibnamefont {Simbotin}},
  \ and\ \bibinfo {author} {\bibfnamefont {R.}~\bibnamefont {C\^ot\'e}},\
  }\href {\doibase 10.1103/PhysRevLett.88.133004} {\bibfield  {journal}
  {\bibinfo  {journal} {Phys. Rev. Lett.}\ }\textbf {\bibinfo {volume} {88}},\
  \bibinfo {pages} {133004} (\bibinfo {year} {2002})}\BibitemShut {NoStop}%
\bibitem [{\citenamefont {Hollerith}\ \emph {et~al.}(2019)\citenamefont
  {Hollerith}, \citenamefont {Zeiher}, \citenamefont {Rui}, \citenamefont
  {Rubio-Abadal}, \citenamefont {Walther}, \citenamefont {Pohl}, \citenamefont
  {Stamper-Kurn}, \citenamefont {Bloch},\ and\ \citenamefont
  {Gross}}]{Hollerith2019}%
  \BibitemOpen
  \bibfield  {author} {\bibinfo {author} {\bibfnamefont {S.}~\bibnamefont
  {Hollerith}}, \bibinfo {author} {\bibfnamefont {J.}~\bibnamefont {Zeiher}},
  \bibinfo {author} {\bibfnamefont {J.}~\bibnamefont {Rui}}, \bibinfo {author}
  {\bibfnamefont {A.}~\bibnamefont {Rubio-Abadal}}, \bibinfo {author}
  {\bibfnamefont {V.}~\bibnamefont {Walther}}, \bibinfo {author} {\bibfnamefont
  {T.}~\bibnamefont {Pohl}}, \bibinfo {author} {\bibfnamefont {D.~M.}\
  \bibnamefont {Stamper-Kurn}}, \bibinfo {author} {\bibfnamefont
  {I.}~\bibnamefont {Bloch}}, \ and\ \bibinfo {author} {\bibfnamefont
  {C.}~\bibnamefont {Gross}},\ }\href@noop {} {\bibfield  {journal} {\bibinfo
  {journal} {Science}\ }\textbf {\bibinfo {volume} {364}},\ \bibinfo {pages}
  {664} (\bibinfo {year} {2019})}\BibitemShut {NoStop}%
\bibitem [{\citenamefont {Hollerith}\ \emph {et~al.}(2021)\citenamefont
  {Hollerith}, \citenamefont {Rui}, \citenamefont {Rubio-Abadal}, \citenamefont
  {Srakaew}, \citenamefont {Wei}, \citenamefont {Zeiher}, \citenamefont
  {Gross},\ and\ \citenamefont {Bloch}}]{Hollerith2021}%
  \BibitemOpen
  \bibfield  {author} {\bibinfo {author} {\bibfnamefont {S.}~\bibnamefont
  {Hollerith}}, \bibinfo {author} {\bibfnamefont {J.}~\bibnamefont {Rui}},
  \bibinfo {author} {\bibfnamefont {A.}~\bibnamefont {Rubio-Abadal}}, \bibinfo
  {author} {\bibfnamefont {K.}~\bibnamefont {Srakaew}}, \bibinfo {author}
  {\bibfnamefont {D.}~\bibnamefont {Wei}}, \bibinfo {author} {\bibfnamefont
  {J.}~\bibnamefont {Zeiher}}, \bibinfo {author} {\bibfnamefont
  {C.}~\bibnamefont {Gross}}, \ and\ \bibinfo {author} {\bibfnamefont
  {I.}~\bibnamefont {Bloch}},\ }\href {\doibase
  10.1103/PhysRevResearch.3.013252} {\bibfield  {journal} {\bibinfo  {journal}
  {Phys. Rev. Research}\ }\textbf {\bibinfo {volume} {3}},\ \bibinfo {pages}
  {013252} (\bibinfo {year} {2021})}\BibitemShut {NoStop}%
\bibitem [{\citenamefont {Niederpr{\"u}m}\ \emph
  {et~al.}(2016{\natexlab{a}})\citenamefont {Niederpr{\"u}m}, \citenamefont
  {Thomas}, \citenamefont {Eichert}, \citenamefont {Lippe}, \citenamefont
  {P{\'e}rez-R{\'\i}os}, \citenamefont {Greene},\ and\ \citenamefont
  {Ott}}]{Niederpruem2016a}%
  \BibitemOpen
  \bibfield  {author} {\bibinfo {author} {\bibfnamefont {T.}~\bibnamefont
  {Niederpr{\"u}m}}, \bibinfo {author} {\bibfnamefont {O.}~\bibnamefont
  {Thomas}}, \bibinfo {author} {\bibfnamefont {T.}~\bibnamefont {Eichert}},
  \bibinfo {author} {\bibfnamefont {C.}~\bibnamefont {Lippe}}, \bibinfo
  {author} {\bibfnamefont {J.}~\bibnamefont {P{\'e}rez-R{\'\i}os}}, \bibinfo
  {author} {\bibfnamefont {C.~H.}\ \bibnamefont {Greene}}, \ and\ \bibinfo
  {author} {\bibfnamefont {H.}~\bibnamefont {Ott}},\ }\href@noop {} {\bibfield
  {journal} {\bibinfo  {journal} {Nature communications}\ }\textbf {\bibinfo
  {volume} {7}},\ \bibinfo {pages} {1} (\bibinfo {year}
  {2016}{\natexlab{a}})}\BibitemShut {NoStop}%
\bibitem [{\citenamefont {Niederpr{\"u}m}\ \emph
  {et~al.}(2016{\natexlab{b}})\citenamefont {Niederpr{\"u}m}, \citenamefont
  {Thomas}, \citenamefont {Eichert},\ and\ \citenamefont
  {Ott}}]{Niederpruem2016b}%
  \BibitemOpen
  \bibfield  {author} {\bibinfo {author} {\bibfnamefont {T.}~\bibnamefont
  {Niederpr{\"u}m}}, \bibinfo {author} {\bibfnamefont {O.}~\bibnamefont
  {Thomas}}, \bibinfo {author} {\bibfnamefont {T.}~\bibnamefont {Eichert}}, \
  and\ \bibinfo {author} {\bibfnamefont {H.}~\bibnamefont {Ott}},\ }\href@noop
  {} {\bibfield  {journal} {\bibinfo  {journal} {Physical Review Letters}\
  }\textbf {\bibinfo {volume} {117}},\ \bibinfo {pages} {123002} (\bibinfo
  {year} {2016}{\natexlab{b}})}\BibitemShut {NoStop}%
\bibitem [{\citenamefont {Duspayev}\ \emph {et~al.}(2021)\citenamefont
  {Duspayev}, \citenamefont {Han}, \citenamefont {Viray}, \citenamefont {Ma},
  \citenamefont {Zhao},\ and\ \citenamefont {Raithel}}]{Duspayev2021}%
  \BibitemOpen
  \bibfield  {author} {\bibinfo {author} {\bibfnamefont {A.}~\bibnamefont
  {Duspayev}}, \bibinfo {author} {\bibfnamefont {X.}~\bibnamefont {Han}},
  \bibinfo {author} {\bibfnamefont {M.}~\bibnamefont {Viray}}, \bibinfo
  {author} {\bibfnamefont {L.}~\bibnamefont {Ma}}, \bibinfo {author}
  {\bibfnamefont {J.}~\bibnamefont {Zhao}}, \ and\ \bibinfo {author}
  {\bibfnamefont {G.}~\bibnamefont {Raithel}},\ }\href@noop {} {\bibfield
  {journal} {\bibinfo  {journal} {Physical Review Research}\ }\textbf {\bibinfo
  {volume} {3}},\ \bibinfo {pages} {023114} (\bibinfo {year}
  {2021})}\BibitemShut {NoStop}%
\bibitem [{\citenamefont {Dei{\ss}}\ \emph {et~al.}(2021)\citenamefont
  {Dei{\ss}}, \citenamefont {Haze},\ and\ \citenamefont
  {Hecker~Denschlag}}]{Deiss2021}%
  \BibitemOpen
  \bibfield  {author} {\bibinfo {author} {\bibfnamefont {M.}~\bibnamefont
  {Dei{\ss}}}, \bibinfo {author} {\bibfnamefont {S.}~\bibnamefont {Haze}}, \
  and\ \bibinfo {author} {\bibfnamefont {J.}~\bibnamefont {Hecker~Denschlag}},\
  }\href@noop {} {\bibfield  {journal} {\bibinfo  {journal} {Atoms}\ }\textbf
  {\bibinfo {volume} {9}},\ \bibinfo {pages} {34} (\bibinfo {year}
  {2021})}\BibitemShut {NoStop}%
\bibitem [{\citenamefont {Zuber}\ \emph {et~al.}(2022)\citenamefont {Zuber},
  \citenamefont {Anasuri}, \citenamefont {Berngruber}, \citenamefont {Zou},
  \citenamefont {Meinert}, \citenamefont {L{\"o}w},\ and\ \citenamefont
  {Pfau}}]{Zuber2022}%
  \BibitemOpen
  \bibfield  {author} {\bibinfo {author} {\bibfnamefont {N.}~\bibnamefont
  {Zuber}}, \bibinfo {author} {\bibfnamefont {V.~S.}\ \bibnamefont {Anasuri}},
  \bibinfo {author} {\bibfnamefont {M.}~\bibnamefont {Berngruber}}, \bibinfo
  {author} {\bibfnamefont {Y.-Q.}\ \bibnamefont {Zou}}, \bibinfo {author}
  {\bibfnamefont {F.}~\bibnamefont {Meinert}}, \bibinfo {author} {\bibfnamefont
  {R.}~\bibnamefont {L{\"o}w}}, \ and\ \bibinfo {author} {\bibfnamefont
  {T.}~\bibnamefont {Pfau}},\ }\href@noop {} {\bibfield  {journal} {\bibinfo
  {journal} {Nature}\ }\textbf {\bibinfo {volume} {605}},\ \bibinfo {pages}
  {453} (\bibinfo {year} {2022})}\BibitemShut {NoStop}%
\bibitem [{\citenamefont {Li}\ \emph {et~al.}(2005)\citenamefont {Li},
  \citenamefont {Tanner},\ and\ \citenamefont {Gallagher}}]{Li2005}%
  \BibitemOpen
  \bibfield  {author} {\bibinfo {author} {\bibfnamefont {W.}~\bibnamefont
  {Li}}, \bibinfo {author} {\bibfnamefont {P.~J.}\ \bibnamefont {Tanner}}, \
  and\ \bibinfo {author} {\bibfnamefont {T.~F.}\ \bibnamefont {Gallagher}},\
  }\href {\doibase 10.1103/PhysRevLett.94.173001} {\bibfield  {journal}
  {\bibinfo  {journal} {Phys. Rev. Lett.}\ }\textbf {\bibinfo {volume} {94}},\
  \bibinfo {pages} {173001} (\bibinfo {year} {2005})}\BibitemShut {NoStop}%
\bibitem [{\citenamefont {Ravets}\ \emph {et~al.}(2014)\citenamefont {Ravets},
  \citenamefont {Labuhn}, \citenamefont {Barredo}, \citenamefont {B{\'e}guin},
  \citenamefont {Lahaye},\ and\ \citenamefont {Browaeys}}]{Ravets2014}%
  \BibitemOpen
  \bibfield  {author} {\bibinfo {author} {\bibfnamefont {S.}~\bibnamefont
  {Ravets}}, \bibinfo {author} {\bibfnamefont {H.}~\bibnamefont {Labuhn}},
  \bibinfo {author} {\bibfnamefont {D.}~\bibnamefont {Barredo}}, \bibinfo
  {author} {\bibfnamefont {L.}~\bibnamefont {B{\'e}guin}}, \bibinfo {author}
  {\bibfnamefont {T.}~\bibnamefont {Lahaye}}, \ and\ \bibinfo {author}
  {\bibfnamefont {A.}~\bibnamefont {Browaeys}},\ }\href@noop {} {\bibfield
  {journal} {\bibinfo  {journal} {Nature Physics}\ }\textbf {\bibinfo {volume}
  {10}},\ \bibinfo {pages} {914} (\bibinfo {year} {2014})}\BibitemShut
  {NoStop}%
\bibitem [{\citenamefont {Holloway}\ \emph {et~al.}(2014)\citenamefont
  {Holloway}, \citenamefont {Gordon}, \citenamefont {Jefferts}, \citenamefont
  {Schwarzkopf}, \citenamefont {Anderson}, \citenamefont {Miller},
  \citenamefont {Thaicharoen},\ and\ \citenamefont {Raithel}}]{Holloway2014}%
  \BibitemOpen
  \bibfield  {author} {\bibinfo {author} {\bibfnamefont {C.~L.}\ \bibnamefont
  {Holloway}}, \bibinfo {author} {\bibfnamefont {J.~A.}\ \bibnamefont
  {Gordon}}, \bibinfo {author} {\bibfnamefont {S.}~\bibnamefont {Jefferts}},
  \bibinfo {author} {\bibfnamefont {A.}~\bibnamefont {Schwarzkopf}}, \bibinfo
  {author} {\bibfnamefont {D.~A.}\ \bibnamefont {Anderson}}, \bibinfo {author}
  {\bibfnamefont {S.~A.}\ \bibnamefont {Miller}}, \bibinfo {author}
  {\bibfnamefont {N.}~\bibnamefont {Thaicharoen}}, \ and\ \bibinfo {author}
  {\bibfnamefont {G.}~\bibnamefont {Raithel}},\ }\href@noop {} {\bibfield
  {journal} {\bibinfo  {journal} {IEEE Transactions on Antennas and
  Propagation}\ }\textbf {\bibinfo {volume} {62}},\ \bibinfo {pages} {6169}
  (\bibinfo {year} {2014})}\BibitemShut {NoStop}%
\bibitem [{\citenamefont {Anderson}\ \emph {et~al.}(2022)\citenamefont
  {Anderson}, \citenamefont {Sapiro}, \citenamefont
  {Gon\ifmmode~\mbox{\c{c}}\else \c{c}\fi{}alves}, \citenamefont {Cardman},\
  and\ \citenamefont {Raithel}}]{Anderson2022}%
  \BibitemOpen
  \bibfield  {author} {\bibinfo {author} {\bibfnamefont {D.}~\bibnamefont
  {Anderson}}, \bibinfo {author} {\bibfnamefont {R.}~\bibnamefont {Sapiro}},
  \bibinfo {author} {\bibfnamefont {L.}~\bibnamefont
  {Gon\ifmmode~\mbox{\c{c}}\else \c{c}\fi{}alves}}, \bibinfo {author}
  {\bibfnamefont {R.}~\bibnamefont {Cardman}}, \ and\ \bibinfo {author}
  {\bibfnamefont {G.}~\bibnamefont {Raithel}},\ }\href {\doibase
  10.1103/PhysRevApplied.17.044020} {\bibfield  {journal} {\bibinfo  {journal}
  {Phys. Rev. Applied}\ }\textbf {\bibinfo {volume} {17}},\ \bibinfo {pages}
  {044020} (\bibinfo {year} {2022})}\BibitemShut {NoStop}%
\bibitem [{\citenamefont {Ripka}\ \emph {et~al.}(2022)\citenamefont {Ripka},
  \citenamefont {Lui}, \citenamefont {Schmidt}, \citenamefont {Kubler},\ and\
  \citenamefont {Shaffer}}]{Ripka2022}%
  \BibitemOpen
  \bibfield  {author} {\bibinfo {author} {\bibfnamefont {F.}~\bibnamefont
  {Ripka}}, \bibinfo {author} {\bibfnamefont {C.}~\bibnamefont {Lui}}, \bibinfo
  {author} {\bibfnamefont {M.}~\bibnamefont {Schmidt}}, \bibinfo {author}
  {\bibfnamefont {H.}~\bibnamefont {Kubler}}, \ and\ \bibinfo {author}
  {\bibfnamefont {J.~P.}\ \bibnamefont {Shaffer}},\ }in\ \href@noop {} {\emph
  {\bibinfo {booktitle} {Optical and Quantum Sensing and Precision Metrology
  II}}},\ Vol.\ \bibinfo {volume} {12016}\ (\bibinfo {organization} {SPIE},\
  \bibinfo {year} {2022})\ pp.\ \bibinfo {pages} {102--107}\BibitemShut
  {NoStop}%
\bibitem [{\citenamefont {Belin}\ and\ \citenamefont
  {Svanberg}(1974)}]{Belin1974}%
  \BibitemOpen
  \bibfield  {author} {\bibinfo {author} {\bibfnamefont {G.}~\bibnamefont
  {Belin}}\ and\ \bibinfo {author} {\bibfnamefont {S.}~\bibnamefont
  {Svanberg}},\ }\href@noop {} {\bibfield  {journal} {\bibinfo  {journal}
  {Physics Letters A}\ }\textbf {\bibinfo {volume} {47}},\ \bibinfo {pages} {5}
  (\bibinfo {year} {1974})}\BibitemShut {NoStop}%
\bibitem [{\citenamefont {Belin}\ \emph
  {et~al.}(1976{\natexlab{a}})\citenamefont {Belin}, \citenamefont {Holmgren},\
  and\ \citenamefont {Svanberg}}]{Belin1976a}%
  \BibitemOpen
  \bibfield  {author} {\bibinfo {author} {\bibfnamefont {G.}~\bibnamefont
  {Belin}}, \bibinfo {author} {\bibfnamefont {L.}~\bibnamefont {Holmgren}}, \
  and\ \bibinfo {author} {\bibfnamefont {S.}~\bibnamefont {Svanberg}},\
  }\href@noop {} {\bibfield  {journal} {\bibinfo  {journal} {Physica Scripta}\
  }\textbf {\bibinfo {volume} {13}},\ \bibinfo {pages} {351} (\bibinfo {year}
  {1976}{\natexlab{a}})}\BibitemShut {NoStop}%
\bibitem [{\citenamefont {Belin}\ \emph
  {et~al.}(1976{\natexlab{b}})\citenamefont {Belin}, \citenamefont {Holmgren},\
  and\ \citenamefont {Svanberg}}]{Belin1976b}%
  \BibitemOpen
  \bibfield  {author} {\bibinfo {author} {\bibfnamefont {G.}~\bibnamefont
  {Belin}}, \bibinfo {author} {\bibfnamefont {L.}~\bibnamefont {Holmgren}}, \
  and\ \bibinfo {author} {\bibfnamefont {S.}~\bibnamefont {Svanberg}},\ }\href
  {\doibase 10.1088/0031-8949/14/1-2/008} {\bibfield  {journal} {\bibinfo
  {journal} {Physica Scripta}\ }\textbf {\bibinfo {volume} {14}},\ \bibinfo
  {pages} {39} (\bibinfo {year} {1976}{\natexlab{b}})}\BibitemShut {NoStop}%
\bibitem [{\citenamefont {Farley}\ \emph {et~al.}(1977)\citenamefont {Farley},
  \citenamefont {Tsekeris},\ and\ \citenamefont {Gupta}}]{Farley1977}%
  \BibitemOpen
  \bibfield  {author} {\bibinfo {author} {\bibfnamefont {J.}~\bibnamefont
  {Farley}}, \bibinfo {author} {\bibfnamefont {P.}~\bibnamefont {Tsekeris}}, \
  and\ \bibinfo {author} {\bibfnamefont {R.}~\bibnamefont {Gupta}},\ }\href
  {\doibase 10.1103/PhysRevA.15.1530} {\bibfield  {journal} {\bibinfo
  {journal} {Phys. Rev. A}\ }\textbf {\bibinfo {volume} {15}},\ \bibinfo
  {pages} {1530} (\bibinfo {year} {1977})}\BibitemShut {NoStop}%
\bibitem [{\citenamefont {Li}\ \emph {et~al.}(2003)\citenamefont {Li},
  \citenamefont {Mourachko}, \citenamefont {Noel},\ and\ \citenamefont
  {Gallagher}}]{Li2003}%
  \BibitemOpen
  \bibfield  {author} {\bibinfo {author} {\bibfnamefont {W.}~\bibnamefont
  {Li}}, \bibinfo {author} {\bibfnamefont {I.}~\bibnamefont {Mourachko}},
  \bibinfo {author} {\bibfnamefont {M.}~\bibnamefont {Noel}}, \ and\ \bibinfo
  {author} {\bibfnamefont {T.}~\bibnamefont {Gallagher}},\ }\href@noop {}
  {\bibfield  {journal} {\bibinfo  {journal} {Physical Review A}\ }\textbf
  {\bibinfo {volume} {67}},\ \bibinfo {pages} {052502} (\bibinfo {year}
  {2003})}\BibitemShut {NoStop}%
\bibitem [{\citenamefont {Steck}(2008)}]{SteckRb}%
  \BibitemOpen
  \bibfield  {author} {\bibinfo {author} {\bibfnamefont {D.~A.}\ \bibnamefont
  {Steck}},\ }\href@noop {} {\enquote {\bibinfo {title} {Rubidium 85 {D} line
  data},}\ }\bibinfo {howpublished}
  {\url{https://steck.us/alkalidata/rubidium85numbers.pdf}} (\bibinfo {year}
  {2008})\BibitemShut {NoStop}%
\bibitem [{\citenamefont {Gallagher}(2005)}]{GallagherBook}%
  \BibitemOpen
  \bibfield  {author} {\bibinfo {author} {\bibfnamefont {T.~F.}\ \bibnamefont
  {Gallagher}},\ }\href@noop {} {\emph {\bibinfo {title} {Rydberg Atoms}}},\
  Vol.~\bibinfo {volume} {3}\ (\bibinfo  {publisher} {Cambridge University
  Press},\ \bibinfo {year} {2005})\BibitemShut {NoStop}%
\bibitem [{\citenamefont {Dieckmann}\ \emph {et~al.}(1998)\citenamefont
  {Dieckmann}, \citenamefont {Spreeuw}, \citenamefont {Weidem{\"u}ller},\ and\
  \citenamefont {Walraven}}]{Dieckmann1998}%
  \BibitemOpen
  \bibfield  {author} {\bibinfo {author} {\bibfnamefont {K.}~\bibnamefont
  {Dieckmann}}, \bibinfo {author} {\bibfnamefont {R.}~\bibnamefont {Spreeuw}},
  \bibinfo {author} {\bibfnamefont {M.}~\bibnamefont {Weidem{\"u}ller}}, \ and\
  \bibinfo {author} {\bibfnamefont {J.}~\bibnamefont {Walraven}},\ }\href@noop
  {} {\bibfield  {journal} {\bibinfo  {journal} {Physical Review A}\ }\textbf
  {\bibinfo {volume} {58}},\ \bibinfo {pages} {3891} (\bibinfo {year}
  {1998})}\BibitemShut {NoStop}%
\bibitem [{\citenamefont {Dalibard}\ and\ \citenamefont
  {Cohen-Tannoudji}(1989)}]{Dalibard1989}%
  \BibitemOpen
  \bibfield  {author} {\bibinfo {author} {\bibfnamefont {J.}~\bibnamefont
  {Dalibard}}\ and\ \bibinfo {author} {\bibfnamefont {C.}~\bibnamefont
  {Cohen-Tannoudji}},\ }\href {\doibase 10.1364/JOSAB.6.002023} {\bibfield
  {journal} {\bibinfo  {journal} {J. Opt. Soc. Am. B}\ }\textbf {\bibinfo
  {volume} {6}},\ \bibinfo {pages} {2023} (\bibinfo {year} {1989})}\BibitemShut
  {NoStop}%
\bibitem [{\citenamefont {Ramos}\ \emph {et~al.}(2019)\citenamefont {Ramos},
  \citenamefont {Cardman},\ and\ \citenamefont {Raithel}}]{Ramos2019}%
  \BibitemOpen
  \bibfield  {author} {\bibinfo {author} {\bibfnamefont {A.}~\bibnamefont
  {Ramos}}, \bibinfo {author} {\bibfnamefont {R.}~\bibnamefont {Cardman}}, \
  and\ \bibinfo {author} {\bibfnamefont {G.}~\bibnamefont {Raithel}},\
  }\href@noop {} {\bibfield  {journal} {\bibinfo  {journal} {Physical Review
  A}\ }\textbf {\bibinfo {volume} {100}},\ \bibinfo {pages} {062515} (\bibinfo
  {year} {2019})}\BibitemShut {NoStop}%
\bibitem [{\citenamefont {Han}\ \emph {et~al.}(2018)\citenamefont {Han},
  \citenamefont {Bai}, \citenamefont {Jiao}, \citenamefont {Hao}, \citenamefont
  {Xue}, \citenamefont {Zhao}, \citenamefont {Jia},\ and\ \citenamefont
  {Raithel}}]{Han2018}%
  \BibitemOpen
  \bibfield  {author} {\bibinfo {author} {\bibfnamefont {X.}~\bibnamefont
  {Han}}, \bibinfo {author} {\bibfnamefont {S.}~\bibnamefont {Bai}}, \bibinfo
  {author} {\bibfnamefont {Y.}~\bibnamefont {Jiao}}, \bibinfo {author}
  {\bibfnamefont {L.}~\bibnamefont {Hao}}, \bibinfo {author} {\bibfnamefont
  {Y.}~\bibnamefont {Xue}}, \bibinfo {author} {\bibfnamefont {J.}~\bibnamefont
  {Zhao}}, \bibinfo {author} {\bibfnamefont {S.}~\bibnamefont {Jia}}, \ and\
  \bibinfo {author} {\bibfnamefont {G.}~\bibnamefont {Raithel}},\ }\href
  {\doibase 10.1103/PhysRevA.97.031403} {\bibfield  {journal} {\bibinfo
  {journal} {Phys. Rev. A}\ }\textbf {\bibinfo {volume} {97}},\ \bibinfo
  {pages} {031403} (\bibinfo {year} {2018})}\BibitemShut {NoStop}%
\bibitem [{\citenamefont {Khazali}(2022)}]{Khazali2022}%
  \BibitemOpen
  \bibfield  {author} {\bibinfo {author} {\bibfnamefont {M.}~\bibnamefont
  {Khazali}},\ }\href@noop {} {\bibfield  {journal} {\bibinfo  {journal}
  {Quantum}\ }\textbf {\bibinfo {volume} {6}},\ \bibinfo {pages} {664}
  (\bibinfo {year} {2022})}\BibitemShut {NoStop}%
\end{thebibliography}%
\end{document}